\documentclass[12pt,a4paper,english]{article}
\usepackage[T1]{fontenc}
\usepackage[latin1]{inputenc}
\usepackage{babel}
\usepackage{graphics}
\usepackage{setspace}
\makeatletter

\providecommand{\LyX}{L\kern-.1667em\lower.25em\hbox{Y}\kern-.125emX\@}

\usepackage{amsfonts}

\makeatother

\begin{document}

\title{Group analysis of the membrane shape equation}
\author{G. De Matteis \footnote{e-mail: dematt73@yahoo.it, giovanni.dematteis@le.infn.it} \\
Dipartimento di Fisica dell'Universit\`a di Lecce, \\ Via Arnesano, CP. 193, 
I-73100 LECCE (Italy).}

\maketitle
\begin{abstract}
A system of partial differential equations describing bilayers amphiphilic membranes
is studied by the Lie group analysis. This algorithmic approach allows us to
show all the symmetries of the system, to determine all possible symmetry reductions,
to recover the axisymmetric solutions, obtained by other euristic methods, and
finally to address the question of new similarity solutions.
\end{abstract}

\section{Introduction}

In the last decades a great interest has been attracted by the membraneous systems
in biophysics \cite{1},\cite{2}. In particular the binary systems amphiphile-water
has been considered \cite{3}. In water the amphiphilic molecules form sheets
of costant area, in the form of bilayers, because this minimizes the contact
of the hydrocarbon tails with water. For a flat piece of membrane, the tails
are still exposed to water contact at the membranes edge. This leads to an effective
line tension which, for large enough membranes, pulls the edges together and
closes the membrane, i.e. we get a \emph{vesicle.} As reviewed in Sec. 2, a
detailed study of such kind of membranes can be performed in a geometrical setting,
following the Helfrich approach. However most of the known analytical solutions
\emph{}\cite{4} can be obtained only under particular geometrical restrictions.
So the aim of this article is to exploit a systematic group analysis (Sec. 3),
in order to find the symmetry properties of the system involving the shape equation
and the Gauss-Codazzi-Mainardi equations. Then, in Sec. 4 we use the above analysis
in performing all the possible symmetry reductions and to determine classes
of particular solutions. Finally in Sec. 5 we address some comments and final
remarks.

\section{The Helfrich functional and the vesicles shape equations}

The membrane's shape is described by a three-component vector field \( \mathbf{x}(u^{1},u^{2})=(x^{1}(u^{1},u^{2}),x^{2}(u^{1},u^{2}),x^{3}(u^{1},u^{2})) \)
depending on the two internal coordinates \( (u^{1},u^{2}) \) of the membrane.
To be self-consistent, we introduce a bit of the geometrical concepts used below
\cite{6},\cite{7},\cite{8}. The tangent plane to a point \( P \) on the
surface is spanned by two tangent vectors \[
\mathbf{t}_{\alpha }=\partial _{\alpha }\mathbf{x},\quad \alpha =1,2\]

where \( \partial _{\alpha }=\frac{\partial }{\partial u^{\alpha }} \).

The two tangent vectors determine the metric tensor or the first fundamental
form \( \Omega _{1} \)\[
g_{\alpha \beta }(u^{1},u^{2})\equiv \mathbf{t}_{\alpha }\cdot \mathbf{t}_{\beta }=\partial _{\alpha }\mathbf{x}\cdot \partial _{\beta }\mathbf{x},\]
 (summation over repeated indices is understood), by which we can express the
infinitesimal element of distance \( ds \) on the surface \[
ds^{2}=g_{\alpha \beta }du^{\alpha }du^{\beta }.\]
 The infinitesimal element of area \( dS \) is given by \[
dS=\sqrt{\det g_{\alpha \beta }}du^{1}du^{2}.\]

The metric tensor \( g_{\alpha \beta } \) characterizes the intrinsic geometry
of the surface. But for a complete description of its embedding into euclidean
three-dimensional space we need the second fundamental form \[
\Omega _{2}=b_{\alpha \beta }du^{\alpha }du^{\beta }\quad \alpha ,\beta =1,2,\]
 where \[
b_{\alpha \beta }(u^{1},u^{2})=(\partial _{\alpha }\partial _{\beta }\mathbf{x})\cdot \mathbf{n}\]
 and \[
\mathbf{n}=\frac{\mathbf{t}_{1}\times \mathbf{t}_{2}}{\left\Vert \mathbf{t}_{1}\times \mathbf{t}_{2}\right\Vert }\]
 is the unit normal vector on the surface. Embedding of the surface into \( \mathbb {R}^{3} \)
is dictated by the Gauss-Weingarten equations (GWE) \begin{equation}
\label{gaussweingar}
\left\{ \begin{array}{c}
\partial _{\alpha }\mathbf{n}=-g^{\beta \gamma }b_{\alpha \gamma }\partial _{\beta }\mathbf{x}\\
\partial _{\alpha }\partial _{\beta }\mathbf{x}=b_{\alpha \beta }\mathbf{n}+\Gamma _{\alpha \beta }^{\sigma }\partial _{\sigma }\mathbf{x}
\end{array}\quad (\alpha ,\beta ,\gamma =1,2),\right. 
\end{equation}
 where \( g^{\beta \gamma }=(g^{-1})_{\beta \gamma } \) and \( \Gamma _{\alpha \beta }^{\sigma } \)
are the so-called Christoffel symbols \[
\Gamma _{\alpha \beta }^{\sigma }=\frac{1}{2}g^{\sigma \delta }\left( \frac{\partial g_{\alpha \delta }}{\partial u^{\beta }}+\frac{\partial g_{\beta \delta }}{\partial u^{\alpha }}-\frac{\partial g_{\alpha \beta }}{\partial u^{\delta }}\right) .\]
 The two most important scalar quantities on the surface are the gaussian curvature
\( K \) and the mean curvature \( H \), given by the determinant and the half-trace
of the matrix \( b_{\alpha }^{\beta }=g^{\beta \gamma }b_{\alpha \gamma }: \)\begin{eqnarray}
K & = & \det (b_{\alpha }^{\beta })=\frac{1}{R_{1}R_{2}},\label{gcurvature} \\
H & = & \frac{1}{2}tr(b_{\alpha }^{\beta })=\frac{1}{2}\left( \frac{1}{R_{1}}+\frac{1}{R_{2}}\right) ,\label{hcurvature} 
\end{eqnarray}
 where \( R_{1} \) and \( R_{2} \) are the principal radii of curvature and
they define the extrinsic properties of the surface. 

Inspired by the physics of liquid crystals, Helfrich \cite{5} considered the
derivatives of the unit vector \( \mathbf{n}\equiv (n^{1},n^{2},n^{3}) \) with
respect to \( u^{1} \) and \( u^{2} \) as the curvature strains. The length
and energy scales are not the size of the molecular head group, or the molecular
energy involved in assembling the amphiphilic sheet. Indeed the basic energy
is that one needed to bend the sheet, i.e. an energy which arises microscopically
from a change in the local concentration of head and tail groups of the amphiphiles.

Over short distances, this energy will ensure that the membrane is relatively
flat. However, over larger distances, the effect of thermal fluctuations is
to change appreciably the orientation of the membrane. The scale of distances
at which an orientation correlation occurs is called the persistence lenght.
It can be calculated from the free energy describing a system of sheets.

We assume that the free energy which describes the deformations of the membrane
depends on its shape only. The local free energy per unit area is written as
an expansion in powers of local curvatures, and the integrated free energy depends
parametrically on the unknown coefficients in this expansion. These coefficients
are the spontaneus, or natural, curvature of the membrane, and its bending and
saddle-splay moduli. Moreover, the free energy must be invariant under translations
and rotations and under reparametrization transformations of the local coordinates,
because of the fluidity assumption. Thus, only the invariant linear and quadratic
forms can enter into, namely \[
\frac{\partial n^{1}}{\partial u^{1}}+\frac{\partial n^{2}}{\partial u^{2}},\quad \left( \frac{\partial n^{1}}{\partial u^{1}}+\frac{\partial n^{2}}{\partial u^{2}}\right) ^{2},\quad \frac{\partial n^{1}}{\partial u^{1}}\frac{\partial n^{2}}{\partial u^{2}}-\frac{\partial n^{1}}{\partial u^{2}}\frac{\partial n^{2}}{\partial u^{1}}.\]
 Then, the elastic free energy takes the form \[
F_{c}=\frac{k}{4}\int _{S}\left( \frac{\partial n^{1}}{\partial u^{1}}+\frac{\partial n^{2}}{\partial u^{2}}-c_{0}\right) ^{2}dA+\overline{k}\int _{S}\left( \frac{\partial n^{1}}{\partial u^{1}}\frac{\partial n^{2}}{\partial u^{2}}-\frac{\partial n^{1}}{\partial u^{2}}\frac{\partial n^{2}}{\partial u^{1}}\right) dA,\]
 where the so-called bending rigidity \( k \) and the gaussian rigidity \( \overline{k} \)
are the elastic moduli. Making use of the GWE (\ref{gaussweingar}) and of the
expressions (\ref{gcurvature}-\ref{hcurvature}) for \( K \) and \( H \),
\( F_{c} \) can be written as \[
F_{c}=k\int _{S}(H-H_{s})^{2}dS+\overline{k}\int _{S}KdS\]
 where \( H_{s}=-\frac{c_{0}}{2} \) is called spontaneus curvature and takes
account of the asimmetry effect of the membrane in its enviroment.

When the bending rigidity \( k\gg k_{B}T \), the persistence length is much
larger than the diameter of a vesicle. Thus we can neglect thermal fluctuations
and the shape of a vesicle is determined by minimizing the elastic energy. Moreover,
the area \( S \) of the vesicle is fixed for two reasons: a) there is very
little exchange of molecules between the bilayer and the ambient on experimental
time scales; b) the elastic energy associated with displacement within the membrane
surface is negligible on the scale of the bending elastic energy. Furthermore,
we can assume that the variations of volume, that is of the water enclosed in,
are negligible. On the contrary, a net transfer of water through the membrane
would lead to a variation of the osmotic pressure, associated with an exchange
of the energy in the system, which is huge with respect to the bending energy
scales. With these costraints, the equilibrium shape of a vesicle is determined
by the minimization of the Helfrich's shape energy \( F \) \cite{1} given
by \begin{equation}
\label{freenergy}
F=k\oint _{S}(H-H_{s})^{2}dS+\overline{k}\oint _{S}KdS+\lambda \oint _{S}dS+\Delta p\int dV
\end{equation}
 where \( \lambda  \) and \( \Delta p=p_{e}-p_{i} \) are treated as Lagrange
multipliers or they can be prescribed experimentally by volume or area reservoirs.
The membrane shapes of equilibrium are defined by the corresponding Euler-Lagrange
equation (the shape equation) \[
\delta ^{(1)}F=0,\]
 where \( \delta ^{(1)} \) indicates the first variation of the functional
under normal infinitesimal distortion of the surface. The resulting equation
is \begin{equation}
\label{shapequation}
\Delta p-2\lambda H-2k(H-H_{s})^{2}H+2k(H-H_{s})(2H^{2}-K)+k\Delta H=0,
\end{equation}
 where \( \Delta  \) applied to \( H \) represents the Laplace-Beltrami operator
on the surface \[
\Delta =\frac{1}{\sqrt{\det g_{ij}}}\partial _{\alpha }\left( g^{\alpha \beta }\sqrt{\det g_{ij}}\partial _{\beta }\right) .\]
 The integrability conditions of the GWE are known from the differential geometry
\cite{6} and they are \begin{equation}
\label{compatiequa}
\begin{array}{c}
\left( \Gamma ^{2}_{12}\right) _{u^{1}}-\left( \Gamma ^{2}_{11}\right) _{u^{2}}+\Gamma ^{1}_{12}\Gamma ^{2}_{11}+\Gamma ^{2}_{12}\Gamma ^{2}_{12}-\Gamma ^{2}_{11}\Gamma ^{2}_{22}-\Gamma ^{1}_{11}\Gamma ^{2}_{12}=-g_{11}K,\\
\frac{\partial b_{11}}{\partial u^{2}}-\frac{\partial b_{12}}{\partial u^{1}}=b_{11}\Gamma ^{1}_{12}+b_{12}(\Gamma ^{1}_{12}-\Gamma ^{1}_{11})-b_{22}\Gamma ^{2}_{11},\\
\frac{\partial b_{12}}{\partial u^{2}}-\frac{\partial b_{22}}{\partial u^{1}}=b_{11}\Gamma ^{1}_{22}+b_{12}(\Gamma ^{2}_{22}-\Gamma ^{1}_{12})-b_{22}\Gamma ^{2}_{12}.
\end{array}
\end{equation}
 These relations are necessary and sufficient conditions for the coefficients
\( g_{\alpha \beta } \) e \( b_{\alpha \beta } \) to be the first and the
second fundamental forms of a surface immersed in the euclidean space. For this
reason they must be included in the set of equations, which describe the shape
of a vesicle. So the equations (\ref{shapequation}) and (\ref{compatiequa})
describe the vesicle configurations in \( \mathbb {R}^{3} \).

In order to simplify the structure of the system, with no loss of generality,
we choose metric on the surface in the conformal form \[
ds^{2}=e^{2\alpha (x,y)}(dx^{2}+dy^{2})\]
 where \( (x,y) \) are conformal local coordinates on the surface. Furthermore,
in order to put in evidence the role of the mean curvature, it is more usefull
to introduce the following variables \begin{eqnarray}
\varphi  & = & \frac{1}{\frac{1}{2}e^{-2\alpha }(b_{11}+b_{22})-H_{s}}=\frac{1}{H-H_{s}},\label{fivariable} \\
q & = & \frac{1}{4}(b_{11}+b_{22})e^{-\alpha }-\frac{H_{s}}{2}e^{\alpha }.\label{qvariable} 
\end{eqnarray}
 Thus, the conformal metric takes the form \[
ds^{2}=4q^{2}\varphi ^{2}(dx^{2}+dy^{2}).\]
 For sake of clarity, we rename the coefficients \( b_{11} \) and \( b_{12} \)
of the second form as \[
b_{11}\equiv \vartheta ,\quad b_{12}\equiv \omega ,\]
 so the associated matrix is \[
\left( \begin{array}{cc}
b_{11} & b_{12}\\
b_{21} & b_{22}
\end{array}\right) =\left( \begin{array}{cc}
\vartheta  & \omega \\
\omega  & 8q^{2}\varphi (1+H_{s}\varphi )-\vartheta 
\end{array}\right) .\]
 In conclusion, the system (\ref{compatiequa}-\ref{shapequation}) takes the
form \begin{equation}
\label{sistemaforma}
\left\{ \begin{array}{c}
q^{2}(\varphi _{xx}+\varphi _{yy})+2q\varphi (q_{xx}+q_{yy})-2\varphi (q_{x}^{2}+q_{y}^{2})+q^{4}(8\varphi +\alpha _{2}\varphi ^{2}+\alpha _{3}\varphi ^{3}+\alpha _{4}\varphi ^{4})=0,\\
\vartheta _{y}-\omega _{x}=(8+\frac{\alpha _{2}}{3}\varphi )q(\varphi q_{y}+q\varphi _{y}),\\
\omega _{y}+\vartheta _{x}=\frac{\alpha _{2}}{3}q\varphi (\varphi q_{x}+q\varphi _{x})+8q\varphi q_{x},\\
4q\varphi ^{2}(q_{xx}+q_{yy})+4\varphi q^{2}(\varphi _{xx}+\varphi _{yy})-4\varphi ^{2}(q_{x}^{2}+q_{y}^{2})-4q^{2}(\varphi _{x}^{2}+\varphi _{y}^{2})=\\
=\omega ^{2}+\vartheta ^{2}-(8+\frac{\alpha _{2}}{3}\varphi )q^{2}\varphi \vartheta ,
\end{array}\right. 
\end{equation}
 where \( \alpha _{2}=24H_{s} \), \( \alpha _{3}=8(2H_{s}^{2}-\frac{\lambda }{k}), \)
\( \alpha _{4}=4\frac{\Delta p}{k}-8H_{s}\frac{\lambda }{k}. \)

This is a system of nonlinear partial differential equations in two dimensions
and in four dependent variables. Its integrability properties are unknown: only
some particular solutions have been found on the basis of euristic considerations
(for example by imposing axisymmetry) \cite{helfrich1},\cite{helfrich2},\cite{4}.

\section{Group analysis}

\subsection{Symmetry algebra}

A specific purpose of the present article is to study the Lie-point symmetry
group of the system (\ref{sistemaforma}). That is, we are looking for a local
Lie group \( G \) of point trasformations on the space of the independent and
dependent variables \[
X\times U=\left\{ \left( x,y,\vartheta ,\omega ,q,\varphi \right) :t.c.:q\neq 0,:\varphi \neq 0\right\} \subseteq \mathbb {R}^{2}\times \mathbb {R}^{4}\]
 into itself, such that each trasformation \( \mathbf{g}\in G \) acts as \[
(\widetilde{x},\widetilde{y},\widetilde{\vartheta },\widetilde{\omega },\widetilde{q},\widetilde{\varphi })=\mathbf{g}\cdot (x,y,\vartheta ,\omega ,q,\varphi )=(\Lambda _{g}^{1},\, \Lambda _{g}^{2},\, \Omega _{g}^{1},\, \Omega _{g}^{2},...,\, \Omega _{g}^{4}),\]
 where \( \Lambda _{g}^{i}=\Lambda _{g}^{i}(x,y,\vartheta ,\omega ,q,\varphi ) \),
\( \Omega _{g}^{j}=\Omega _{g}^{j}(x,y,\vartheta ,\omega ,q,\varphi ) \) are
smooth real functions, depending also on a suitable set of \( r \) real parameters
\( g \), continously varying in a neighbourhood of the origin in \( \mathbb {R}^{r} \).
In this section we will follow the standard group analysis scheme developped
in \cite{Olver},\cite{ovsiannikov},\cite{winternitz}. For a symmetry group
of a system of differential equations we mean that each element \( \mathbf{g} \)
trasforms solutions of the system (\ref{sistemaforma}) into other solutions:
\( (\widetilde{\vartheta }(\widetilde{x},\widetilde{y}),\widetilde{\omega }(\widetilde{x},\widetilde{y}),\widetilde{q}(\widetilde{x},\widetilde{y}),\widetilde{\varphi }(\widetilde{x},\widetilde{y})) \)
is a solution whenever \( ((\vartheta (x,y),\omega (x,y),q(x,y),\varphi (x,y)) \)
is one. Equivalently, we can say that the symmetry group leaves the system invariant.
Moreover we are interested in connected local Lie groups of symmetries, leaving
aside the problem of the discrete symmetries. Instead of considering finite
trasformations, in the study of connected groups, we can restrict our attention
to the associated algebra, that is to the infinitesimal trasformations. The
symmetry algebra \( \widehat{g} \) of the system is realized by real vector
fields, of the form \[
\mathbf{v}=\xi ^{1}\frac{\partial }{\partial x}+\xi ^{2}\frac{\partial }{\partial y}+\phi _{1}\frac{\partial }{\partial \vartheta }+\phi _{2}\frac{\partial }{\partial \omega }+\phi _{3}\frac{\partial }{\partial q}+\phi _{4}\frac{\partial }{\partial \varphi },\]
 where \( \xi ^{i} \) and \( \phi _{\alpha } \) are functions of \( x,y,\vartheta ,\omega ,q,\varphi  \).
These functions have to be determined by the invariance condition of the differential
system (\ref{sistemaforma}), namely \begin{equation}
\label{invarieq}
pr^{(2)}\mathbf{v}\left[ \Delta _{\nu }\right] =0\; \; whenever\; \Delta _{\nu }=0,\quad for\; \nu =1,...,4,
\end{equation}
 where \( \Delta _{\nu }=0 \) indicates each of the equations in (\ref{sistemaforma}),
and \[
pr^{(2)}\mathbf{v}=\mathbf{v}+\sum _{\alpha =1}^{4}\phi _{\alpha }^{x}\frac{\partial }{\partial u_{x}^{\alpha }}+\sum _{\alpha =1}^{4}\phi _{\alpha }^{y}\frac{\partial }{\partial u_{y}^{\alpha }}+\sum _{\alpha =1}^{4}\phi _{\alpha }^{xx}\frac{\partial }{\partial u_{xx}^{\alpha }}+\sum _{\alpha =1}^{4}\phi _{\alpha }^{xy}\frac{\partial }{\partial u_{xy}^{\alpha }}+\sum _{\alpha =1}^{4}\phi _{\alpha }^{yy}\frac{\partial }{\partial u_{yy}^{\alpha }},\]
 is the prolonged vectorfield on the space containing the derivatives of the
dependent variables. Here we have indicated with \( u^{1}=\vartheta ,\, u^{2}=\omega ,\, u^{3}=q,\, u^{4}=\varphi  \)
and the components \( \phi _{\alpha }^{s} \) and \( \phi _{\alpha }^{st} \)
can be computed algorithmically \cite{Olver}. The equations (\ref{invarieq})
lead to a large number of linear partial differential equations, called determining
equations, for the functions \( \xi ^{i} \), \( \phi _{\alpha } \). Solving
them, we can distinguish two cases:

\( I) \) if the phenomenological parameters are not vanishing \( (\alpha _{2},\alpha _{3},\alpha _{4})\neq (0,0,0) \),
the symmetry algebra \( \widehat{L}_{I} \) is spanned by the vector fields
\begin{eqnarray*}
\mathbf{v}(\xi ^{1}(x,y),\xi ^{2}(x,y)) & = & \mathbf{v}(\xi (z))=\xi \partial _{z}+\\
 &  & +\xi _{z}\left[ -(\vartheta -i\omega )\partial _{\vartheta }-(\omega +i\vartheta -4iq^{2}\varphi -i\frac{\alpha _{2}}{6}q^{2}\varphi ^{2})\partial _{\omega }-\frac{q}{2}\partial _{q}\right] +\\
 &  & +c.c.,
\end{eqnarray*}
 where \( z=x+iy \) and \( \xi =\xi ^{1}+i\xi ^{2} \), being \( \xi ^{1},\xi ^{2} \)
arbitrary real harmonic functions satisfying the Cauchy-Riemann conditions \[
\xi _{y}^{1}=-\xi _{x}^{2},\; \xi _{x}^{1}=\xi _{y}^{2};\]

\( II) \) restricting the phenomenological parameters to be \( (\alpha _{2},\alpha _{3},\alpha _{4})=(0,0,0) \),
the symmetry algebra \( \widehat{L}_{II} \) is spanned by the vector fields
\begin{eqnarray*}
\mathbf{v}_{1}(\xi ^{1}(x,y),\xi ^{2}(x,y)) & = & \mathbf{v}_{1}(\xi (z))=\xi \partial _{z}+\\
 &  & +\xi _{z}\left[ -(\vartheta -i\omega )\partial _{\vartheta }-(\omega +i\vartheta -4iq^{2}\varphi )\partial _{\omega }-\frac{q}{2}\partial _{q}\right] +\\
 &  & +c.c.,\\
\mathbf{v}_{2} & = & \vartheta \partial _{\vartheta }+\omega \partial _{\omega }+\varphi \partial _{\varphi },
\end{eqnarray*}
 with the same notation used above. The non trivial commutation relations for
\( \widehat{L}_{I} \) are \begin{equation}
\label{commu1}
\left[ \mathbf{v}(\xi (z)),\mathbf{v}(\widetilde{\xi }(z))\right] =\mathbf{v}(\xi \widetilde{\xi }_{z}-\widetilde{\xi }\xi _{z}),
\end{equation}
 and for \( \widehat{L}_{II} \)\begin{eqnarray}
\left[ \mathbf{v}_{1}(\xi (z)),\mathbf{v}_{1}(\widetilde{\xi }(z))\right] =\mathbf{v}_{1}(\xi \widetilde{\xi }_{z}-\widetilde{\xi }\xi _{z}), &  & \label{comm1} \\
\left[ \mathbf{v}_{1}(\xi (z)),\mathbf{v}_{2}\right] =0, &  & \label{comm2} 
\end{eqnarray}
 where \( \xi (z) \) and \( \widetilde{\xi }(z) \) are arbitrary complex analitic
functions. In both cases the symmetry Lie algebras are infinite -dimensional.
In particular, the algebra \( \widehat{L}_{I} \) is a realization of the conformal
transformations algebra in the plane \( \widehat{L}_{c} \) \cite{18}. Moreover,
\( \widehat{L}_{II} \) is the direct sum of \( \widehat{L}_{c} \) and of the
one-dimensional subalgebra \( \widehat{L}_{g} \) generated by \( \mathbf{v}_{2} \)\[
\widehat{L}_{II}=\widehat{L}_{c}\oplus \widehat{L}_{g}.\]
 A more explicit base for the algebra \( \widehat{L}_{I} \) \( (\widehat{L}_{c}) \)
is obtainable by developing the functions \( \xi (z) \) in Laurent series,
so that the generic vector field is \begin{eqnarray*}
L_{n}(\alpha ) & = & e^{i\alpha }z^{n+1}\partial _{z}+\\
 & + & e^{i\alpha }(n+1)z^{n}\left[ -(\vartheta -i\omega )\partial _{\vartheta }-(\omega +i\vartheta -4iq^{2}\varphi -i\frac{\alpha _{2}}{6}q^{2}\varphi ^{2})\partial _{\omega }-\frac{q}{2}\partial _{q}\right] +\\
 &  & +c.c.,
\end{eqnarray*}
 where \( \alpha \in \left[ 0,2\pi \right[  \). Thus the commutation relations
become \begin{equation}
\label{commutators}
\left[ L_{n}(\alpha ),L_{m}(\beta )\right] =(m-n)L_{n+m}(\alpha +\beta ).
\end{equation}
 In passing, we note that for \( \alpha =0 \) the generators \( L_{n}(0) \)
span a centerless Virasoro algebra \cite{19}. The generators \( \left\{ L_{-1}(0),L_{-1}(\frac{\pi }{2}),L_{0}(0),L_{0}(\frac{\pi }{2}),L_{1}(0),L_{1}(\frac{\pi }{2})\right\}  \)
span the \( sl(2,\mathbb {C}) \) algebra, which is the algebra of Möbius transformations
in the complex plane of local coordinates of the surface.

In order to give an interpretation of this symmetry structure of the system
(\ref{sistemaforma}), we consider the effect of finite transformations on the
metric. The finite transformations generated by vector fields \( \mathbf{v} \)
and \( \mathbf{v}_{1} \)on the variables \( q \), \( \varphi  \), \( z \),
\( \overline{z} \) are given by \begin{equation}
\label{transfindvar}
\widetilde{z}(\varepsilon ,z)=\Gamma ^{-1}(\varepsilon +\Gamma (z)),\quad \overline{\widetilde{z}}(\varepsilon ,\overline{z})=\overline{\Gamma ^{-1}}(\varepsilon +\overline{\Gamma (\overline{z})}),
\end{equation}
 where \( \varepsilon  \) is the group parameter and \[
\Gamma (z)=\int _{z_{0}}^{z}\frac{ds}{\xi (s)}.\]
 While \( q \) and \( \varphi  \) transform as \begin{equation}
\label{qfitransform}
\widetilde{q}(\lambda ,q,z,\overline{z})=q(z,\overline{z})\left| \frac{\xi (z)}{\xi (\widetilde{z})}\right| ,\quad \widetilde{\varphi }(\varphi ,z,\overline{z})=\varphi .
\end{equation}
 Moreover, being \[
\frac{\partial z}{\partial \widetilde{z}}=\frac{\xi (z(\widetilde{z}))}{\xi (\widetilde{z})},\]
 we have \[
d\widetilde{s}^{2}=4\widetilde{q}^{2}\widetilde{\varphi }^{2}d\widetilde{z}d\overline{\widetilde{z}}=4q^{2}(z(\widetilde{z}),\overline{z}(\overline{\widetilde{z}}))\varphi ^{2}(z(\widetilde{z}),\overline{z}(\overline{\widetilde{z}}))dzd\overline{z}=ds^{2},\]
 that is the metric is invariant under group transformations. Moreover, since
the mean curvature \( H \) is related to \( \varphi  \) via the eq. (\ref{fivariable}),
by virtue of the eq. (\ref{qfitransform}), we find that \( H \) remains invariant
or, equivalently, it transforms like a scalar under the group transformations.
Concerning the gaussian curvature \( K \), it is a metric invariant, thus it
remains invariant under the symmetry group because the metric tensor does. In
conclusion we have seen that both \( H \) and \( K \) are invariant under
the class obtained of transformations. Consequently, a surface, which is a solution
of system (\ref{sistemaforma}), cannot be {}``deformed{}'' into another one
by using a group transformation.

\subsection{Classification of the low dimensional subalgebras}

In order to perform all possible symmetry reductions of (\ref{sistemaforma}),
we must proceed to the subalgebras classification under the inner automorphism
group (or adjoint group). This leads to build up the so-called optimal system,
that is the set of the representatives of the \( s \)-dimensional subalgebra
classes, which are pairwise nonconjugated. We start by considering the conjugation
classes of \( 1 \)-dimensional subalgebras. A first simple result can be achieved
by computing the adjoint action on a generic element \( L_{m}(\alpha ) \) (with
\( m\neq 0 \)) from the commutation relations (\ref{commutators}) \[
Ad\left[ \exp (\varepsilon L_{0}(\alpha ))\right] \left\langle L_{m}(\beta )\right\rangle =e^{-m\varepsilon \cos \alpha }L_{m}(\beta -m\varepsilon \sin \alpha ),\]
 with \( \varepsilon \in \mathbb {R} \), and \( \alpha ,\beta \in \left[ 0,2\pi \right[  \).
Then, by a suitable choice of the parameters, we can conjugate any subalgebra
spanned by \( \left\{ L_{m}(\beta )\right\} _{m\neq 0} \) to \( \left\{ L_{m}(0)\right\} _{m\neq 0} \).
More generally, this treatment can be applied looking for the action of a generic
element \( \mathbf{v}(\xi ) \) on a vector field \( \mathbf{v}(\delta ) \)
(being \( \xi =\xi ^{1}+i\xi ^{2} \) and \( \delta =\delta ^{1}+i\delta ^{2} \)
analytic functions of \( z=x+iy \)) \cite{kpeq}. Integrating the corresponding
differential equation we are lead to the functional equation \[
\delta (\widetilde{z})=\mu (z(\widetilde{z}))\frac{\xi (\widetilde{z})}{\xi (z(\widetilde{z}))},\]
 where \( z(\widetilde{z}) \) can be obtained from (\ref{transfindvar}) and
\[
\delta (\widetilde{z})\left| _{\varepsilon =0}=\right. \mu (z)\]
 for \( \mu (z) \) given. Now, if we would like to conjugate \( \mathbf{v}(\mu (z)) \)
to the translation generator \( \mathbf{v}(1) \), we have to find a transformation
generated by a vectorfield \( \mathbf{v}(\xi ), \) by solving the functional
equation for \( \xi  \): \[
1=\mu (z(\widetilde{z}))\frac{\xi (\widetilde{z})}{\xi (z(\widetilde{z}))}.\]
 Now, the rectification theorem for vectorfields \cite{arnold} suggests and
Neumann proved \cite{neumann} that the real version of this functional equation
admits a solution for smooth \( \mu  \) and out of a neighbourhood of its zeros.
Thus, we can conclude that all \( 1 \)-dimensional subalgebras of \( \widehat{L}_{I} \)
are conjugated among themselves, at least in a suitable domain. Therefore, we
can choose the translation vectorfield \( L_{-1}(0)=\mathbf{v}(1) \) as the
representative of all \( 1 \)-dimensional subalgebras. We now consider a \( 2 \)-dimensional
subalgebra \( \left\{ \mathbf{v}(\zeta ),\mathbf{v}(\eta )\right\}  \). In
view of the above result, we can conjugate \( \mathbf{v}(\zeta ) \) to \( \mathbf{v}(1) \)
and \( \mathbf{v}(\eta ) \) will be transformed into some other \( \mathbf{v}(\widetilde{\eta }). \)
In addition the commutation relations (\ref{commu1}-\ref{comm2}) and the closure
condition require \[
\left[ \mathbf{v}(1),\mathbf{v}(\widetilde{\eta })\right] =\mathbf{v}(\widetilde{\eta }_{z})=a\mathbf{v}(1)+b\mathbf{v}(\widetilde{\eta }),\]
 that is \[
\widetilde{\eta }(z)=-\frac{a}{b}+ce^{bz},\quad c\in \mathbb {C},c\neq 0.\]
 Performing the transformation \( \widetilde{z}=e^{-bz}, \) we end up with
\[
\partial _{z}=-b\widetilde{z}\partial _{\widetilde{z}},\quad e^{bz}\partial _{z}=-b\partial _{\widetilde{z}}.\]
 Thus, we can conclude that all the \( 2 \)-dimensional subalgebras are conjugated
among themselves and, precisely, to the subalgebra \[
\widehat{L}_{td}=\left\{ L_{-1}(0),L_{0}(0)\right\} ,\]
 which can be taken as a representative of all \( 2 \)-dimensional subalgebras.

\subsection{Reduced systems}

In order to find explicit group invariant solutions it is enough to restrict
ourselves to \( 1 \)- and \( 2 \)-dimensional subalgebras. In fact, in these
cases we have \( 4 \) group invariants, which are in one- to- one corrispondence
with the dependent variables, so that their role can be interchanged. Concerning
the independent variables, we have respectively \( 1 \) and \( 0 \) new indipendent
similarity variables. Thus, the original partial differential system is reduced
to an ordinary differential one or to a functional one, respectively.

We now perform the symmetry reductions of (\ref{sistemaforma}) starting from
the \( 2 \)-dimensional subalgebras, by considering the representative subalgebra
\( \widehat{L}_{td} \). The correspondent subgroup has the following basic
invariants \begin{eqnarray*}
I_{1}=s\vartheta ^{1/2}, &  & I_{2}=s\omega ^{1/2},\\
I_{3}=sq, &  & I_{1}=\varphi ,
\end{eqnarray*}
 where \( s=z-\overline{z} \). From which we obtain \begin{eqnarray*}
\varphi =\varphi _{0}, &  & q=\frac{I_{3}}{s}=\frac{Q}{y},\\
\vartheta =\frac{I_{1}^{2}}{s^{2}}=\frac{\Theta }{y^{2}}, &  & \vartheta =\frac{I_{2}^{2}}{s^{2}}=\frac{\Omega }{y^{2}},
\end{eqnarray*}
 where \( \Theta ,\Omega ,Q,\varphi _{0} \) are arbitrary real costants. By
substituting these relations in (\ref{sistemaforma}) we obtain \[
\begin{array}{c}
Q(\varphi _{0})=-\frac{2}{8+\alpha _{2}\varphi _{0}+\alpha _{3}\varphi _{0}^{2}+\alpha _{4}\varphi _{0}^{3}},\\
\Theta (\varphi _{0})=\frac{1}{2}\varphi _{0}Q^{2}\left( 8+\frac{\alpha _{2}}{3}\varphi _{0}\right) ,\\
\Omega =0,\\
16=-Q^{2}\left( 8+\frac{\alpha _{2}}{3}\varphi _{0}\right) ^{2}.
\end{array}\]
 The inconsistency of the last equation induces us to conclude that there are
not group invariant solutions with respect to any \( 2 \)-dimensional subgroup.

Now we consider solutions invariant under \( 1 \)-dimensional subgroups, all
represented by the translation group generated by the algebra \( \left\{ L_{-1}(0)\right\}  \).
With respect to the previous case, we have to add only one more invariant, namely
\( s \), that is \( y. \) Then, the invariant solutions have the form \[
\vartheta =\vartheta (y),\; \omega =\omega (y),\; q=q(y),\; \varphi =\varphi (y).\]
 The reduced system becomes a system of ordinary differential equations \begin{equation}
\label{unidimsystem}
\begin{array}{c}
q^{2}\frac{d^{2}\varphi }{dy^{2}}+2q\varphi \frac{d^{2}q}{dy^{2}}-2\varphi \left( \frac{dq}{dy}\right) ^{2}+q^{4}\left( 8\varphi +\alpha _{2}\varphi ^{2}+\alpha _{3}\varphi ^{3}+\alpha _{4}\varphi ^{4}\right) =0,\\
\frac{d\vartheta }{dy}=\left( 8+\frac{\alpha _{2}}{3}\varphi \right) q\left( \varphi \frac{dq}{dy}+q\frac{d\varphi }{dy}\right) ,\\
4q\varphi ^{2}\frac{d^{2}q}{dy^{2}}+4\varphi q^{2}\frac{d^{2}\varphi }{dy^{2}}-4\varphi ^{2}\left( \frac{dq}{dy}\right) ^{2}-4q^{2}\left( \frac{d\varphi }{dy}\right) ^{2}=\\
=\alpha _{5}^{2}+\vartheta ^{2}-\left( 8+\frac{\alpha _{2}}{3}\varphi \right) q^{2}\varphi \vartheta ,
\end{array}
\end{equation}
 where we put \( \omega =\alpha _{5} \) being a constant. In the next section
we will look for some particular solutions of this system.

The importance of this system lies in the fact that all possible \( 1 \)-dimensional
symmetry reductions lead to it, modulo transformations of the dependent and
independent variables. In particular, this has to be true for all systems obtained
by using some clever parametrization of the axysymmetric surfaces studied in
the litterature \cite{helfrich1},\cite{helfrich2},\cite{4}.

\section{Solutions}

In order to find particular solutions of (\ref{unidimsystem}), we make some
simplifying hypotheses. The aim is to give explicit solutions to the reduced
system for the amphiphilic membranes configurations.

\subsection{Sphere}

We first consider the case \( \varphi =\varphi _{0} \). This condition means
that we are taking under consideration only the constant mean curvature surfaces,
since \begin{equation}
\label{ficost}
\varphi _{0}=\frac{1}{H_{0}-H_{s}},\quad H_{0}\neq H_{s}.
\end{equation}
 The system (\ref{unidimsystem}) takes the form \begin{equation}
\label{ficostsist}
\begin{array}{c}
2q\varphi _{0}\frac{d^{2}q}{dy^{2}}-2\varphi _{0}\left( \frac{dq}{dy}\right) ^{2}+q^{4}\left( 8\varphi _{0}+\alpha _{2}\varphi _{0}^{2}+\alpha _{3}\varphi _{0}^{3}+\alpha _{4}\varphi _{0}^{4}\right) =0,\\
\frac{d\vartheta }{dy}=\left( 8+\frac{\alpha _{2}}{3}\varphi _{0}\right) q\varphi _{0}\frac{dq}{dy},\\
4q\varphi _{0}^{2}\frac{d^{2}q}{dy^{2}}-4\varphi _{0}^{2}\left( \frac{dq}{dy}\right) ^{2}=\alpha _{5}^{2}+\vartheta ^{2}-\left( 8+\frac{\alpha _{2}}{3}\varphi _{0}\right) q^{2}\varphi _{0}\vartheta .
\end{array}
\end{equation}
 The second equation gives \[
\vartheta (y)=\frac{B(\varphi _{0})}{2}q^{2}+\vartheta _{0},\]
 where \( B(\varphi _{0})=\left( 8+\frac{\alpha _{2}}{3}\varphi _{0}\right) \varphi _{0}. \)
The remaining equations give \begin{equation}
\label{sphericalshapeeq}
q\frac{d^{2}q}{dy^{2}}-\left( \frac{dq}{dy}\right) ^{2}+q^{4}\frac{A(\varphi _{0})}{2\varphi _{0}}=0,
\end{equation}
 where \( A(\varphi _{0})=\left( 8\varphi _{0}+\alpha _{2}\varphi _{0}^{2}+\alpha _{3}\varphi _{0}^{3}+\alpha _{4}\varphi _{0}^{4}\right)  \),
with the conditions \[
\begin{array}{c}
\alpha _{5}=0,\\
\vartheta _{0}=0,\\
B^{2}(\varphi _{0})=8\varphi _{0}A(\varphi _{0}).
\end{array}\]
 Assuming \( A(\varphi _{0})\neq 0, \) the last relation is a condition on
the mean curvature in terms of the phenomenological parameters, namely \begin{equation}
\label{helfrichcond}
2H_{s}kH_{0}^{2}-2H_{s}^{2}kH_{0}-2\lambda H_{0}+\Delta p=0.
\end{equation}
 A non trivial solution (modulo translations) of (\ref{sphericalshapeeq}) is
\[
q(y)=\frac{\sqrt{\delta _{0}}}{C}\frac{1}{\cosh (\sqrt{\delta _{0}}y)},\]
 where \( C=2\left| H_{0}\varphi _{0}\right|  \) and \( \delta _{0} \) is
a constant of integration. The gaussian curvature is \[
K=-\frac{1}{4q^{2}\varphi ^{2}}\nabla ^{2}\ln (q\varphi )=H_{0}^{2}.\]
 Thus, the described surface is a sphere of radius \( R=\frac{1}{H_{0}} \)
under the condition (\ref{helfrichcond}). By performing a coordinate change
\( x\prime =\sqrt{\delta _{0}}x,\: y\prime =\sqrt{\delta _{0}}y \), the metric
and the second fundamental form result \[
\begin{array}{c}
ds^{2}=\frac{1}{H_{0}^{2}\cosh ^{2}(y\prime )}(dx\prime ^{2}+dy\prime ^{2}),\\
\Omega _{2}=\frac{1}{H_{0}}\frac{1}{\cosh ^{2}(y\prime )}(dx\prime ^{2}+dy\prime ^{2}),
\end{array}\]
 and the representation in the three-dimensional space is given by \[
x^{1}(x\prime ,y\prime )=-\frac{R\sin x\prime }{\cosh y\prime },\; x^{2}(x\prime ,y\prime )=-\frac{R\cos x\prime }{\cosh y\prime },\; x^{3}(x\prime ,y\prime )=-R\tanh y\prime .\]

\subsection{Delaunay's surfaces}

\begin{figure}
{\par\centering \resizebox*{5cm}{!}{\includegraphics{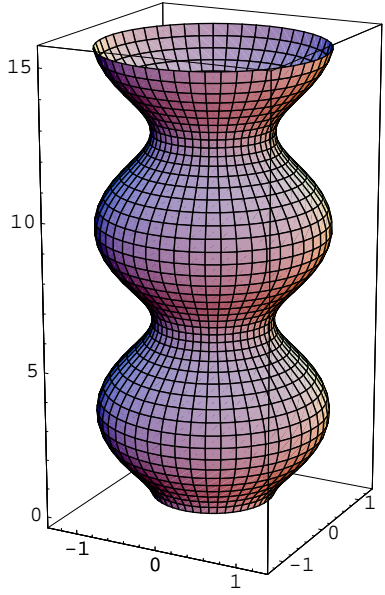}} \par}

\caption{Unduloid \protect\( \beta _{1}=0.5\protect \), \protect\( \vartheta _{0}=7.2\protect \)\label{unduloid}}
\end{figure}

\begin{figure}
{\par\centering \resizebox*{5cm}{!}{\includegraphics{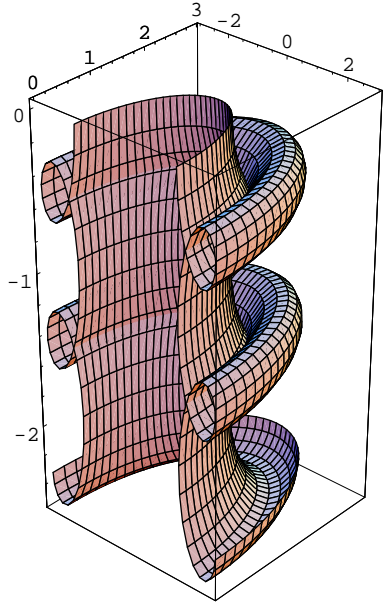}} \par}

\caption{Nodoid \protect\( \beta _{1}=0.8\protect \) \protect\( \vartheta _{0}=-25\protect \)\label{nodoid1}}
\end{figure}

\begin{figure}
{\par\centering \resizebox*{5cm}{!}{\includegraphics{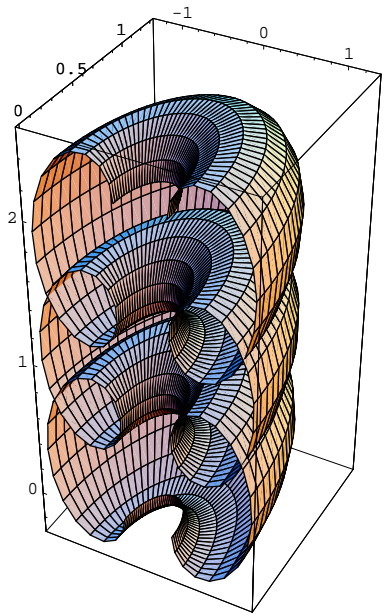}} \par}

\caption{Nodoid \protect\( \beta _{1}=1\protect \) \protect\( \vartheta _{0}=-1.6\protect \)\label{nodoid2}}
\end{figure}

\begin{figure}
{\par\centering \resizebox*{5cm}{!}{\includegraphics{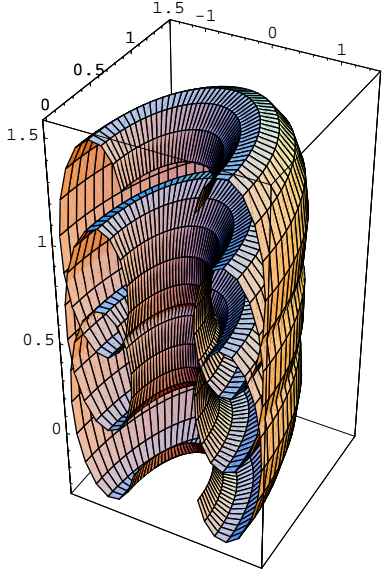}} \par}

\caption{Nodoid \protect\( \beta _{1}=1\protect \) \protect\( \vartheta _{0}=-2.8\protect \)\label{nodoid3}}
\end{figure}

We now consider costant mean curvature surfaces in the particular case of \( H_{0}=H_{s} \).
In this case we cannot use the relation (\ref{ficost}), so that we are forced
to perform some slight changes. We first introduce the following variables \[
\begin{array}{c}
p=q(1+H_{s}\varphi ),\\
\widetilde{\varphi }=\frac{\varphi }{1+H_{s}\varphi }=\frac{1}{H},
\end{array}\]
 in terms of which the reduced system takes the form \begin{equation}
\label{pfitildesystem}
\begin{array}{c}
p^{2}\frac{d^{2}\widetilde{\varphi }}{dy^{2}}-2\beta _{1}p^{2}\left( \widetilde{\varphi }\frac{d^{2}\widetilde{\varphi }}{dy^{2}}-\left( \frac{d\widetilde{\varphi }}{dy}\right) ^{2}\right) +2\widetilde{\varphi }\left( 1-\beta _{1}\widetilde{\varphi }\right) \left( p\frac{d^{2}p}{dy^{2}}-\left( \frac{dp}{dy}\right) ^{2}\right) +\\
+8p^{4}\widetilde{\varphi }-8\beta _{1}^{2}p^{4}\widetilde{\varphi }^{3}+4\beta _{3}p^{4}\widetilde{\varphi }^{4}-8\beta _{2}p^{4}\widetilde{\varphi }^{3}=0,\\
\frac{d\vartheta }{dy}=8p\left( \widetilde{\varphi }\frac{dp}{dy}+p\frac{d\varphi }{dy}\right) ,\\
4\widetilde{\varphi }^{2}p\frac{d^{2}p}{dy^{2}}+4p^{2}\widetilde{\varphi }\frac{d^{2}\widetilde{\varphi }}{dy^{2}}-4p^{2}\left( \frac{d\widetilde{\varphi }}{dy}\right) ^{2}-4\widetilde{\varphi }^{2}\left( \frac{dp}{dy}\right) ^{2}=\vartheta ^{2}+\alpha _{5}^{2}-8p^{2}\widetilde{\varphi }\vartheta ,
\end{array}
\end{equation}
 where \( \beta _{1}=H_{s},\; \beta _{2}=\frac{\lambda }{k},\; \beta _{3}=\frac{\Delta p}{k}. \)
By imposing the condition \( \widetilde{\varphi }=\frac{1}{H_{s}}, \) the first
equation in (\ref{pfitildesystem}) gives \[
\Delta p=2\lambda H_{s}.\]
 The second equation becomes \[
\vartheta (y)=\frac{4}{\beta _{1}}p^{2}(y)+\vartheta _{0},\]
 and the last one takes the form \begin{equation}
\label{odeshapedelaunay}
p\frac{d^{2}p}{dy^{2}}-\left( \frac{dp}{dy}\right) ^{2}=-4p^{4}+D^{2},
\end{equation}
 where \( D^{2}=\frac{1}{4}\beta _{1}^{2}(\vartheta _{0}^{2}+\alpha _{5}^{2}). \)
Equation (\ref{odeshapedelaunay}) is a slight generalization of equation (\ref{sphericalshapeeq}).
The integration of (\ref{odeshapedelaunay}) gives\begin{equation}
\label{taldeitali}
p(y)=\sqrt{r_{2}}\sqrt{1-\sigma ^{2}sn^{2}(2\sqrt{r_{2}}(y-y_{0}),\sigma )}
\end{equation}
where \( r_{2}=\frac{\delta _{0}+\sqrt{\delta ^{2}_{0}-4D^{2}}}{4} \), \( \sigma =\sqrt{\frac{8\sqrt{\delta ^{2}_{0}-4D^{2}}}{\delta _{0}+\sqrt{\delta ^{2}_{0}-4D^{2}}}} \)
and \( \delta _{0}>2\left| D\right|  \)a constant of integration. Notice that
for \( D^{2}=0 \), that is \( \vartheta _{0}=\alpha _{5}=0 \), we obtain the
sphere of radius \( R=\frac{1}{H_{s}} \). Furthermore, choosing \( D=\pm \frac{\delta _{0}}{2}, \)
we get a cylinder with \( H=H_{s} \). In the general case, the two fundamental
forms are\[
\begin{array}{c}
ds^{2}=\frac{4r_{2}}{\beta ^{2}_{1}}\left[ 1-\sigma ^{2}sn^{2}\left( 2\sqrt{r_{2}}(y-y_{0}),\sigma \right) \right] \left( dx^{2}+dy^{2}\right) ,\\
\Omega _{2}=\left[ \frac{4r_{2}}{\beta _{1}}\left[ 1-\sigma ^{2}sn^{2}\left( 2\sqrt{r_{2}}(y-y_{0}),\sigma \right) \right] +\vartheta _{0}\right] dx^{2}+2\alpha _{5}dxdy+\\
+\left[ \frac{4r_{2}}{\beta _{1}}\left[ 1-\sigma ^{2}sn^{2}\left( 2\sqrt{r_{2}}(y-y_{0}),\sigma \right) \right] -\vartheta _{0}\right] dy^{2}.
\end{array}\]
Without loss of generality, we can always choose a coordinate system in which
\( \alpha _{5}=0 \). Indeed both \( H \) and \( K \) depend only on \( \delta _{0} \)
and \( D \). In order to simplify the last expressions, we perform a coordinate
change \( (x,y)\rightarrow (\phi ,p) \) by the equation (\ref{taldeitali})
for \( y \) and via the relation \( x=\frac{\beta _{1}}{2}\phi  \). Thus we
obtain\begin{equation}
\label{delaunaysur}
\begin{array}{c}
ds^{2}=-\frac{1}{\beta ^{2}_{1}}\frac{p^{2}}{p^{4}-\frac{\delta _{0}}{2}p^{2}+\frac{D^{2}}{4}}dp^{2}+p^{2}d\phi ^{2},\\
\Omega _{2}=\frac{1}{\beta _{1}}\frac{p^{2}-\frac{\vartheta _{0}\beta _{1}}{4}}{-p^{4}+\frac{\delta _{0}}{2}p^{2}-\frac{D^{2}}{4}}dp^{2}+\\
+\beta _{1}\left( p^{2}+\frac{\beta _{1}\vartheta _{0}}{4}\right) d\phi ^{2}.
\end{array}
\end{equation}
These two forms represent a family of surfaces parametrized by \( \beta _{1},\vartheta _{0},\delta _{0} \).
In the particular case \( \delta _{0}=\frac{2-\vartheta _{0}\beta ^{3}_{1}}{\beta ^{2}_{1}} \)
the two forms (\ref{delaunaysur}) generate the Delaunay surfaces \cite{23}\cite{22}.
Precisely we have the unduloids when \( 0<\beta ^{3}_{1}\vartheta _{0}<1 \)
(figure\ref{unduloid}), and the nodoids when \( \beta ^{3}_{1}\vartheta _{0}<0 \)
(figures \ref{nodoid1},\ref{nodoid2},\ref{nodoid3}). Otherwise, when the
above constraint on \( \delta _{0} \) is not satisfied, we guess that a larger
class of surfaces appears. However we leave this study to the next future.

\subsection{Toroidal surfaces}

\begin{figure}
{\par\centering \resizebox*{5cm}{!}{\includegraphics{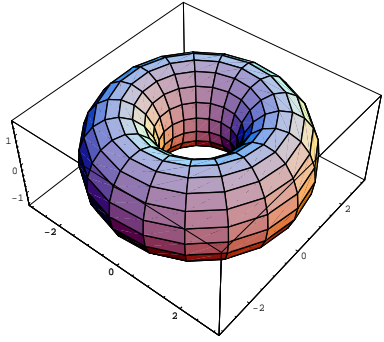}} \par}

\caption{Clifford torus\label{toruus}}
\end{figure}

Now, we leave the class of the constant mean cuevature surfaces and consider
the reduction of the system (\ref{unidimsystem}) under the particular restriction
\( q=q_{0} \) a constant. The reduced system (\ref{unidimsystem}) takes the
form \begin{equation}
\label{qconstsyst}
\begin{array}{l}
\frac{d^{2}\varphi }{dy^{2}}+q_{0}^{2}\left( 8\varphi +\alpha _{2}\varphi ^{2}+\alpha _{3}\varphi ^{3}+\alpha _{4}\varphi ^{4}\right) =0\\
\frac{d\vartheta }{dy}=\left( 8+\frac{\alpha _{2}}{3}\varphi \right) q_{0}^{2}\frac{d\varphi }{dy},\\
4\varphi q_{0}^{2}\frac{d^{2}\varphi }{dy^{2}}-4q_{0}^{2}\left( \frac{d\varphi }{dy}\right) ^{2}=\alpha _{5}^{2}+\vartheta ^{2}-\left( 8+\frac{\alpha _{2}}{3}\varphi \right) q_{0}^{2}\varphi \vartheta .
\end{array}
\end{equation}
 Integrating the second equation we obtain \[
\vartheta (y)=q_{0}^{2}\varphi \left( 8+\frac{\alpha _{2}}{6}\varphi \right) +c_{1},\]
 which substituted in the third equation of (\ref{qconstsyst}) gives \[
\varphi \frac{d^{2}\varphi }{dy^{2}}=\left( \frac{d\varphi }{dy}\right) ^{2}+\frac{\alpha _{5}^{2}+c_{1}^{2}}{4q_{0}^{2}}-\frac{\alpha _{2}^{2}q_{0}^{2}}{144}\varphi ^{4}+2c_{1}\varphi -\frac{\alpha _{2}}{3}q_{0}^{2}\varphi ^{3}.\]
 By comparing this equation with the first one of the system (\ref{qconstsyst}),
we obtain\[
\left( \frac{d\varphi }{dy}\right) ^{2}+\frac{\alpha ^{2}_{5}+c^{2}_{1}}{4q^{2}_{0}}+2c_{1}\varphi +8q^{2}_{0}\varphi ^{2}+\frac{2\alpha _{2}}{3}q^{2}_{0}\varphi ^{3}+q^{2}_{0}\left( \alpha _{3}-\frac{\alpha ^{2}_{2}}{144}\right) \varphi ^{4}+q^{2}_{0}\alpha _{4}\varphi ^{5}=0\]
Furthermore, the integration of the first equation (\ref{qconstsyst}) gives\[
\left( \frac{d\varphi }{dy}\right) ^{2}=-8q^{2}_{0}\varphi ^{2}-\frac{2}{3}q^{2}_{0}\alpha _{2}\varphi ^{3}-\frac{\alpha _{2}}{3}q^{2}_{0}\varphi ^{4}-\frac{2}{5}q^{2}_{0}\alpha _{4}\varphi ^{5}+\phi _{0}.\]
So the comparison of the last two equations provides the constraints\[
\begin{array}{c}
\alpha _{4}=0\Leftrightarrow \Delta p=2\lambda H_{s},\\
\frac{\alpha _{3}}{2}=\frac{\alpha ^{2}_{2}}{144}\Leftrightarrow \lambda =kH^{2}_{s},\\
c_{1}=0,\\
\phi _{0}=-\frac{\alpha ^{2}_{5}}{4q^{2}_{0}}.
\end{array}\]
Consequently the original system is reduced to the equation\begin{equation}
\label{toruseq}
\left( \frac{d\varphi }{dy}\right) ^{2}=-\gamma _{0}-\gamma _{2}\varphi ^{2}-\gamma _{3}\varphi ^{3}-\frac{\gamma ^{2}_{3}}{8\gamma _{2}}\varphi ^{4}=P_{4}(\varphi ),
\end{equation}
where \( \gamma _{0}=\frac{\alpha ^{2}_{5}}{4q^{2}_{0}},\: \gamma _{2}=8q^{2}_{0},\: \gamma _{3}=16H_{s}q^{2}_{0} \).
In order to ensure the positivity of the r.h.s. of (\ref{toruseq}) we must
impose the condition\[
\alpha ^{2}_{5}<\frac{8q^{6}_{0}}{H^{2}_{s}}(11+5\sqrt{5}).\]
As suggested by the equation (\ref{toruseq}), performing the coordinate transformation
\( (x,y)\rightarrow (x,\varphi ) \), the two fundamental forms are\begin{eqnarray*}
ds^{2} & = & \frac{\gamma _{2}}{2}\varphi ^{2}\left( dx^{2}-\frac{d\varphi ^{2}}{\gamma _{0}+\gamma _{2}\varphi ^{2}+\gamma _{3}\varphi ^{3}+\frac{\gamma ^{2}_{3}}{8\gamma _{2}}\varphi ^{4}}\right) ,\\
\Omega _{2} & = & \varphi \left( \gamma _{2}+\frac{\gamma _{3}}{4}\varphi \right) dx^{2}+\frac{2\alpha _{5}}{\sqrt{-\gamma _{0}-\gamma _{2}\varphi ^{2}-\gamma _{3}\varphi ^{3}-\frac{\gamma ^{2}_{3}}{8\gamma _{2}}\varphi ^{4}}}dxd\varphi +\\
 &  & +\frac{\gamma _{3}\varphi ^{2}}{4\left( -\gamma _{0}-\gamma _{2}\varphi ^{2}-\gamma _{3}\varphi ^{3}-\frac{\gamma ^{2}_{3}}{8\gamma _{2}}\varphi ^{4}\right) }d\varphi ^{2},
\end{eqnarray*}
with \( 0<\beta <\varphi <\alpha  \) if \( H_{s}<0 \) and \( \beta <\varphi <\alpha <0 \)
if \( H_{s}>0 \), being \( \alpha ,\beta  \) the two real roots of \( P_{4}(\varphi ), \)
which is positive in \( \left] \alpha ,\beta \right[  \). The difference in
the sign of \( H_{s} \) depends only on the different orientation of the normal
vector \( \mathbf{n} \). The gaussian and mean curvatures are given by\[
H=\frac{1}{\varphi }+\frac{\gamma _{3}}{2\gamma _{2}},\quad K=-\frac{2\gamma _{0}}{\gamma _{2}\varphi ^{4}}+\frac{2\gamma _{3}}{\gamma _{2}\varphi }+\frac{\gamma ^{2}_{3}}{4\gamma ^{2}_{2}}.\]
In the special case \( \alpha _{5}=0 \) and rescaling the coordinates \( x=\frac{1}{\sqrt{\gamma _{2}}}\theta ,\: \varphi =\sqrt{2}\rho , \)
we have\begin{eqnarray}
ds^{2} & = & \rho ^{2}d\theta ^{2}-\frac{d\rho ^{2}}{1+2\sqrt{2}H_{s}\rho +H^{2}_{s}\rho ^{2}},\label{tr} \\
\Omega _{2} & = & \left( H_{s}\rho ^{2}+\sqrt{2}\rho \right) d\theta ^{2}-\frac{H_{s}d\rho ^{2}}{1+2\sqrt{2}H_{s}\rho +H^{2}_{s}\rho ^{2}},\label{tr2} \\
H & = & H_{s}+\frac{\sqrt{2}}{2\rho },\label{tr3} \\
K & = & H^{2}_{s}+\frac{\sqrt{2}H_{s}}{\rho }.\label{torus} 
\end{eqnarray}
If we interpret the symmetry variable \( \theta  \) as the rotation angle through
the \( x^{3} \)-axis and \( \rho  \) as the radius \( \sqrt{(x^{1})^{2}+(x^{2})^{2}} \),
then (\ref{tr}-\ref{tr2}) are the two fundamental forms of the so called Clifford
torus, with cartesian equation\[
\left[ (x^{1})^{2}+(x^{2})^{2}+(x^{3})^{2}+\frac{1}{H^{2}_{s}}\right] ^{2}=\frac{8}{H^{2}_{s}}\left[ (x^{1})^{2}+(x^{2})^{2}\right] .\]
This vesicle configuration has been detected experimentally by M. Mutz and Bensimon
\cite{24},\cite{25}(see figure \ref{toruus}).

\subsection{Circular Biconcave Discoid }

\begin{figure}
{\par\centering \includegraphics{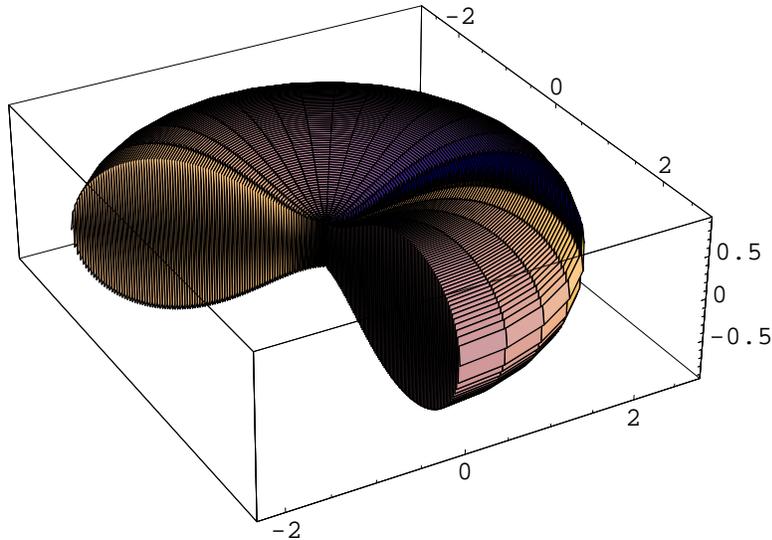} \par}

\caption{Circular Biconcave Surface\label{redblood}}
\end{figure}

In the previous sections we have considered the special cases in which the variables
\( q \) or \( \varphi  \) are taken to be constant. In this section we will
consider the general case in which both \( q \) and \( \varphi  \) are not
constant functions, but we suppose that \( q \) is functionally depending on
\( \varphi  \). Thus we put without loss of generality\begin{equation}
\label{relation1}
q=q(\varphi )=\frac{\rho (\varphi )}{2\varphi },
\end{equation}
where \( \rho  \) is a function to be determined. We now look at the system
(\ref{unidimsystem}) and consider the second equation. Assuming a one-to-one
correspondence between \( \rho  \) and \( \varphi  \), we have\[
\vartheta =\left( \frac{\alpha _{2}}{24}+\frac{1}{\varphi }\right) \rho ^{2}+\int \frac{\rho ^{2}}{\varphi ^{2}}\frac{d\varphi }{d\rho }d\rho +\vartheta _{0},\]
where the integral is still an undetermined function of \( \rho  \). However
one could play with in order to get closed expressions. For example, the toroidal
case can be recovered by assuming \( \varphi  \) and \( \rho  \) to be proportional,
or the expression under integral of the form \( \frac{c}{\rho ^{2}}. \) Another
interesting possibility arises making the following hypothesis\( : \)\begin{equation}
\label{relation2}
\frac{1}{\varphi ^{2}}\frac{d\varphi }{d\rho }=-\frac{c}{\rho },
\end{equation}
or, equivalently\[
\varphi =\frac{1}{c\ln \left( \frac{\rho }{\rho _{0}}\right) }.\]
Consequently we have\[
\vartheta =\left( \frac{\alpha _{2}}{24}-\frac{c}{2}+c\ln \left( \frac{\rho }{\rho _{0}}\right) \right) \rho ^{2}+\vartheta _{0}.\]
If we now substitute the two last expressions in the remaining equations of
(\ref{unidimsystem}), we obtain the compatibility conditions\begin{equation}
\label{condiscoid}
\alpha _{2}=12c,\; \alpha _{3}=4c^{2},\; \alpha _{4}=0,\; \alpha _{5}=\vartheta _{0}=0,
\end{equation}
thus \( c=2H_{s} \) and the other phenomenological parameters \( \lambda  \)
and \( \Delta p \) are equal to zero because of (\ref{condiscoid}). Furthermore,
we have the following differential equation for \( \rho (y) \)\begin{equation}
\label{discoideq}
\frac{d}{dy}\left( \frac{\frac{d\rho }{dy}}{\rho }\right) +c^{2}\rho ^{2}\ln \left( \frac{\rho }{\rho _{0}}\right) \left( 1+\ln \left( \frac{\rho }{\rho _{0}}\right) \right) =0.
\end{equation}
In order to integrate this equation we perform the hodograph transformation\begin{equation}
\label{hodograph}
\frac{dy}{d\rho }=f^{-\frac{1}{2}}(\rho ).
\end{equation}
The equation (\ref{discoideq}) provides a differential equation for \( f(\rho ) \)\[
\frac{df}{d\rho }-\frac{2}{\rho }f=-2c^{2}\rho ^{3}\ln \left( \frac{\rho }{\rho _{0}}\right) \left( 1+\ln \left( \frac{\rho }{\rho _{0}}\right) \right) ,\]
the general solution of which is\[
f(\rho )=\rho ^{2}\left( f_{0}-c^{2}\rho ^{2}\ln ^{2}\left( \frac{\rho }{\rho _{0}}\right) \right) \qquad (f_{0}\in \mathbb {R}).\]
Thus, the problem of the system (\ref{unidimsystem}) with tha assumptions (\ref{relation1}-\ref{relation2})
is reduced to the quadrature of (\ref{hodograph}). Furthermore, we can conclude
that the solution surface is defined by the following two fundamental forms
in the local coordinate \( (x,\rho ) \) \begin{eqnarray*}
ds^{2} & = & \rho ^{2}dx^{2}+\frac{\rho ^{2}}{f(\rho )}d\rho ^{2},\\
\Omega _{2} & = & c\rho ^{2}\ln \left( \frac{\rho }{\rho _{0}}\right) dx^{2}+c\rho ^{2}\left( \ln \left( \frac{\rho }{\rho _{0}}\right) +1\right) \frac{1}{f(\rho )}d\rho ^{2}.
\end{eqnarray*}
From the above expression of the metric, we interpret \( \rho  \) as a radius
and \( x \) as an angular variable, obtaining an axisymmetric surface. A general
parametrization of this kind of surfaces is given in terms of angle \( \psi  \)
between the tangent to the contour and the radial direction \cite{4}\begin{eqnarray*}
x^{1}=\cos (x) &  & x^{2}=\sin (x)\\
 & x^{3}=\int \tan \psi (\rho )d\rho +z_{0} & 
\end{eqnarray*}
where \( \sec ^{2}\psi (\rho )=\frac{\rho ^{2}}{f(\rho )}. \) The required
integration for \( x^{3} \), which can be numerically performed, leads to the
well known \cite{4} biconcave discoid surface (see figure \ref{redblood})
where we set \( f_{0}=1 \) in order to avoid a conical point in \( \rho =0 \).
This surface may represent a static configuration for the red blood cells.

\section{Conclusions}

In this article we have shown that the symmetry algebra of the system (\ref{sistemaforma})
is the algebra of the conformal transformations in two dimensions, that is they
generate arbitrary reparametrizations of the local coordinates \( (x,y) \),
accompained by the appropriate transformation of the dependent variables. In
the particular case \( H_{s}=\Delta p=\lambda =0 \) the functional \( F \)
(\ref{freenergy}) becomes the so-called Willmore functional and the symmetries
include a one-dimensional algebra \( \widehat{L}_{g} \) (spanned by the vectorfield
\( \mathbf{v}_{2} \)), which generates the scaling transformations in the ambient
space \cite{9}\cite{10}. One of the main result of our research is that in
the general case (the most important from the physical point of view) the group
analysis leads to a unique symmetry reduction, that is the system (\ref{unidimsystem}).
By making different simplifying hypothesis (\( q=q_{0} \), \( \varphi =\varphi _{0} \),
etc...) we have recovered the known solutions of toroidal and constant mean
curvature type, respectively. However, the first and second fundamental forms
contain possible generalizations of these surface solutions. Moreover we have
seen how to obtain in a systematic way other classes of solutions, by imposing
particular relations between \( q \) and \( \varphi  \) (see eq. \ref{relation1}-\ref{relation2}).
In this way we have obtained the so-called circular biconcave surfaces, but
we presume to be able to reduce to the quadratures many other classes of surfaces
in the next future. Besides, in order to find new solutions and study their
stability, we plan to use the discrete symmetries of the system, numerical methods
in determining solutions of the GWE (\ref{gaussweingar}) and finally we plan
to study the second variation of the functional \( F \) on the solutions.

\section*{Acknowledgments}

The work presented here was supported by a research grant from the Dipartimento
di Fisica of the Universita' di Lecce and MIUR.

The author thanks prof. L. Martina for many helpfull discussions.

\end{document}